\documentclass{aa}
\usepackage{graphics}
\usepackage{color}
\begin{document}
\thesaurus{12                       % A&A Section 12: Cosmology
           (12.07.1;                % (Cosmology:) gravitational lensing
            12.04.1;                % (Cosmology:) dark matter 
            11.17.4 Q0957+561;      % (Galaxies:) quasars: individual
            03.13.4)}               % Methods: numerical 

\title{Limits on MACHOs from microlensing in the double quasar
Q0957+561}

\author{Robert Schmidt \and Joachim Wambsganss}

\offprints{R. Schmidt, email: rschmidt@aip.de}

\institute{Astrophysikalisches Institut Potsdam, An der Sternwarte 16,
14482 Potsdam, Germany}

\date{\today}

\titlerunning {Limits on MACHOs from quasar microlensing}% in Q0957+561}
\authorrunning {R. Schmidt \& J. Wambsganss}

\maketitle
\begin{abstract} 

The light curves of the two images of the double quasar
\object{Q0957+561} as obtained by Kundi\'c et al. (1997) are almost
identical, except for an overall time delay and scaling factor.  This
allows us to put limits on the amount of microlensing that took place
during the time interval corresponding to the monitoring
observations. We perform numerical simulations in which we model the
microlensing behaviour of the (halo of the) lensing galaxy in the
system.  We test ``MACHO-masses'' ranging from $10^{-8}$ to $10^{-1}$
M$_{\odot}$ and quasar sizes from $10^{14}$ to $3\times 10^{15}$ cm.
Statistically comparing the expected microlensing-induced changes from
100\,000 simulated light curves over a period of 160 days with the
(lack of) observed fluctuations, we can constrain regions in the
parameter space of MACHO mass and quasar size with various degrees of
confidence.  In particular, a halo consisting of objects at the low
end of our mass scale can be ruled out with high confidence for a
small quasar size.  A halo consisting of objects with $10^{-2}$ or
$10^{-1}$ M$_{\odot}$ cannot be ruled out yet, but it should produce
MACHO induced fluctuations in future observations. We also test halos
with only 50\% or 25\% of the mass in compact objects; constraints
here are a bit less stringent.

\end{abstract}


\section{Introduction}

It has been known for almost three decades that rotation curves of
galaxies remain flat even beyond the visible matter (Rubin \& Ford
\cite{Rubin1970}, Rubin et al. \cite{Rubin1985}). As an explanation,
Ostriker et al. (\cite{Ostriker1974}) and Einasto et
al. (\cite{Einasto1974}) suggested that galaxies are surrounded by
large halos consisting of ``dark matter'' and dominating the total
mass of the galaxies. Over the years a very large number of possible
``candidates'' for this dark matter have been suggested (for recent
reviews see, e.g., Bahcall \cite{Bahcall1997} or Raffelt
\cite{Raffelt1997}). They can broadly be divided into ``elementary
particle'' candidates (e.g. massive neutrinos or axions) and
``astrophysical'' candidates (e.g. black holes, brown dwarfs,
``Jupiters'', comets). Not much progress has been made to date in
identifying the elusive dark matter despite major efforts in many
directions. In fact, a new branch of physics established itself --
``astro-particle physics'' -- whose major goal is the solution of the
dark matter problem.

More than 10 years ago Paczy\'nski (\cite{Paczynski1986}) proposed a
direct test to prove or reject the possibility that the halo of the
Milky Way consists of dark compact objects by way of gravitational
microlensing: If the brightness of at least a million stars in the
Large Magellanic Cloud (LMC) could be regularly observed, at any given
time about one of them should be magnified due to a halo object (in
the following referred to as ``MACHO'', for ``massive compact halo
object'') passing in front of it and focussing the light rays to the
observer. The light curve of the affected background star should show
a very characteristic and achromatic behaviour.

Soon thereafter various groups started big observational programs to
investigate this promising possibility, and the first microlensing
events were found in 1993 (Alcock et al. \cite{Alcock1993}; Aubourg et
al. \cite{Aubourg1993}). The most recent results indicate that about
$50^{+30}_{-20}\%$ of the dark matter in the Milky Way halo can
consist of such objects (Alcock et al.  \cite{Alcock1997}). Their most
likely mass range ($M \approx 0.5 ^{+0.3}_{-0.2}$ M$_\odot$), however,
is higher than originally suggested for genuine ``brown dwarfs'', but
the uncertainty is large.  The largest problem in these experiments is
the small probability for a microlensing event: the ``optical depth''
(the fraction of background stars that are significantly affected by
microlensing) is less than $10^{-6}$.

In a very different optical depth regime, gravitational lensing can
also be used to test whether halos of {\it other} galaxies are made of
compact objects (Gott \cite{Gott1981}): In cases of good projected
alignment along the line of sight between a background qua\-sar and a
foreground galaxy, the galaxy can produce multiple images of the
quasar\footnote{Note that
	here the distances involved are about five orders of magnitude
	larger than in the microlensing searches towards the LMC (few
	Gigaparsec rather than some 55 kpc), and the optical depths
	involved are of order unity (i.e. six orders of magnitude
	higher).}.
In this case the light bundles from the quasar pass through (the dark
halo of) the foreground galaxy on their way to the observer. These
light bundles probe the graininess of the halo: if the dark matter
there consists of some kind of elementary particles (``smoothly
distributed matter'' on astronomical length scales), the light bundles
should be unaffected; if it is made of compact astrophysical objects,
the light bundles can probe it.  The measured brightnesses of the
quasar images should vary as a function of time due to the changing
relative positions between lens, source and observer (Chang \& Refsdal
\cite{Chang1979}).  The surface mass density (optical depth) in these
cases is high enough (of order unity) to basically affect (change) the
measured flux of a quasar image all the time.

Since quasars are intrinsically variable as well, it is not trivial to
decide whether an observed variability is intrinsic or
microlensing-induced. In the case of multiple images of one quasar,
however, the intrinsic fluctuations should show up coherently in all
the images -- modulo a certain time delay -- whereas the microlensing
changes occur uncorrelated in the various images. Once well-sampled
light curves of two (or more) quasar images are obtained and the time
delay $\Delta t$ (due to the different light paths of different light
bundles) and the magnitude difference $\Delta m$ (due to different
magnifications) is known (both of which are not trivial), one can shift
the two light curves in time and magnitude by the appropriate amounts
$\Delta t$ and $\Delta m$ and subtract them from each other.  All
remaining fluctuations in the ``difference light curve'' must be due
to microlensing. In particular, if the difference light curve is flat,
this indicates that there was no microlensing going on during the
period of observation\footnote{The effect of ``flat'' microlensing --
periods in which microlensing produces a constant (de-) magnification
-- are taken care of by a different magnification ratio.}.

The gravitationally lensed double quasar Q0957\linebreak[0]+561 (Walsh
et al. \cite{Walsh1979}) has been observed/monitored for almost two
decades by many groups (see e.g., Schild \& Thomson \cite{Schild1995})
with the goal to measure the time delay and subsequently determine the
Hubble constant.  The time delay is now established firmly at $\Delta
t = 417\pm3$ days (Schild \& Thomson \cite{Schild1997}; Kundi\'c et
al. \cite{Kundic1997}, subsequently K97; Oscoz et
al. \cite{Oscoz1997}). In addition, a number of gravitational lens
models have been published for the Q0957+561 system which reproduce
the observed (optical and radio) image configurations well (Falco et
al. \cite{Falco1991}; Grogin \& Narayan \cite{Grogin1996}), so that
the parameters relevant for microlensing at the position of the quasar
images can be determined from these models with an accuracy of a few
percent.

With both the time delay established and good models being available,
the double quasar Q0957+561 can hence be used as a test for massive
halo objects by comparison of the light curves of images A and B.  We
do this here by simulating the microlensing effect numerically for
MACHOs of different masses.  We analyse the resulting microlensed
light curves and show how often microlensing-induced changes of
certain amplitudes are to be expected for certain MACHO masses.
Finally, we compare our numerical simulations with the well-sampled
data set of K97 and thus constrain the MACHO masses in (the halo of)
the lensing galaxy.

In Sect. 2 we describe briefly the monitoring data set of K97 for
the double quasar and the simulations we perform in order to compare
the observations with microlensing light curves due to different
masses.  In Sect. 3 we show our results for a MACHO mass range from
$10^{-7} < M/M_{\odot} < 10^{-1}$, and in the final Sect. 4 we
present our conclusions.

\section{Observations and simulations}
\label{obsandth}
We first briefly present the macro model of the lens system and the
recent data set on the quasar Q0957+561 which we use. We then define
the relevant numbers and parameters and illustrate how the
microlensing light curve depends on the MACHO masses and the quasar
size.  Finally, the numerical technique is described which we employ
to produce the microlensing light curves and to analyse them.

\subsection {Lens model and optical data}
\label{lensmodanddata}
The double quasar Q0957+561 ($z=1.41$) is gravitationally lensed by a
galaxy ($z=0.36$) and its associated galaxy cluster. The complex
structure of the quasar in the radio regime places strong constraints
on theoretical models (e.g. Gorenstein et al.
\cite{Gorenstein1983,Gorenstein1988}), so that well determined values
for the surface mass density and the local shear at the positions of
the quasar images can be obtained (Falco et
al. \cite{Falco1991}, priv. comm., Grogin \& Narayan \cite{Grogin1996}).

The normalized surface mass density $\kappa = \Sigma / \Sigma_{crit}$
is the projected mass density of the lensing galaxy plus cluster along
the line of sight, normalized by the critical surface mass density
(Schneider et al. \cite{Schneider1992})
\begin{equation}
\label{sigmacrit}
\Sigma_{\rm crit}=\frac{c^2}{4\pi G} \frac{D_{\rm s}}{D_{\rm d}D_{\rm
	ds}}=0.92\,h\,{\rm g}/{\rm cm}^{2}.
\end{equation}
Here $D_{\rm d}$, $D_{\rm s}$ and $D_{\rm ds}$ are the angular
diameter distances between observer, deflector and source,
respectively (the velocity of light is denoted by $c$ and the
gravitational constant by $G$). The underlying cosmological model
is an Einstein de-Sitter universe with a Hubble constant of
H$_0=100\,h\,$km$\,$s$^{-1}$Mpc$^{-1}$.

The local shear $\gamma$ represents the tidal field at the image
position due to the matter outside the beam.  We adopt values for the
surface density of $\kappa_A = 0.32$ for image A and $\kappa_B = 1.17$
for image B, for the local shear at the image positions we used
$\gamma_A = 0.18$ and $\gamma_B = 0.83$, respectively (Falco priv. comm.).

Recently, K97 presented the results of two years of well sampled
monitoring observations of Q0957+561 with the Apache Point
Observatory. With their $g$ band data, they confirmed the value of
$417\pm3$ days for the time delay $\Delta t$ between the two lensed
light paths and a best value for the magnitude offset between the
(time-corrected) quasar fluxes of
\begin{equation}
	\Delta m_{AB} := 
	\ \ \ < m_{\rm A}(t)-m_{\rm B}(t+\Delta t) > \ \ \ 
		= 0.118\, {\rm mag}.
\end{equation}

In order to quantitatively estimate the effects or limits of
microlensing on the light curves of Q0957+561 A and/or B, we determined
the difference light curve between the two quasar images in the
following way. First, the flux-corrected and time shifted
light curve of image B (i.e.: $m_{B}(t)\rightarrow m'_{B}(t)= m_{B}
(t+\Delta t)+ 0.118\, {\rm mag} $) was subtracted from the light curve
of image A:
\begin{equation}
\Delta m(t) = m_{A}(t)  - m'_{B}(t),
\end{equation}
which measures the deviation between the two light curves.

\begin{figure}
\resizebox{\hsize}{!}{\includegraphics{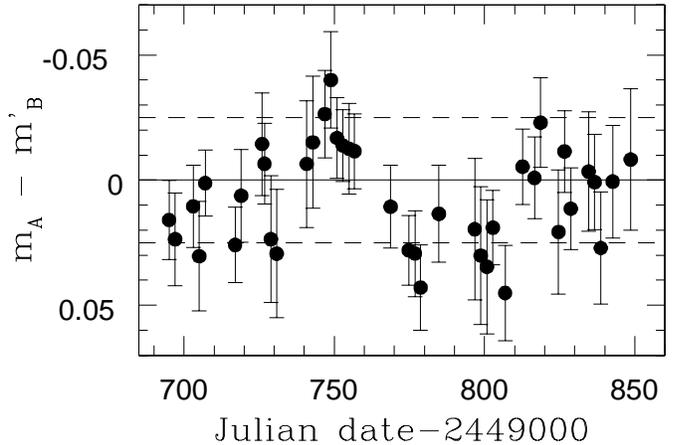}}
\caption{Difference light curve of images A and B (shifted in time and
magnitude) of the quasar Q0957+561. To guide the eye, dashed lines are
drawn at differences of +0.025 mag and -0.025 mag.}
\label{diff}
\end{figure}

Since the light curve had been sampled only for discrete points in
time, we had to interpolate between the two closest points of
light curve B before and after the instant of time in which a point of
light curve A had been determined (or vice versa):
\begin{equation}
m'_{B}(t_j) = c_i\: m'_{B}(t_i) + c_k\: m'_{B}(t_k),
\end{equation}
where $c_i = (t_j - t_i) / (t_k - t_i)$ and $c_i + c_k = 1$ and $t_i <
t_j < t_k$; $t_j$ indicates the time at which data was taken for image
A, and $t_i, t_k$ are the closest times before and after $t_i$ for
which data exists for image B.  The ``difference light curve'' is
defined as:
\begin{equation}
\label{deltam}
\Delta m(t_j) = m_{A}(t_j)  - m'_{B}(t_j).
\end{equation}
The measurement uncertainties $\sigma(t_j)$ were added quadratically
for each combined pair of data during interpolation and
subtraction. The resulting difference light curve is shown in
Fig.~\ref{diff}. We also determined the difference light curve by
interpolating the A-light curve (rather than B) and obtained a very
similar result.

We used 36 observations of image A and 40 observations of image B out
of the 99 available observations from the years 1995 and 1996 (K97)
since we had to restrict ourselves to the about 160 days of
``overlap'' between the image A light curve and the time-shifted image
B light curve.  The time axis in Fig.~\ref{diff} corresponds to the
observing epoch of image A.

We detect no variation in the difference light curve with an amplitude
greater than $\approx 0.05$ mag. And considering the error bars we find
that the difference light curve would even be consistent with $\Delta
m=0$ mag. There is also no systematic gradient apparent in the data,
which would be the signature of a long term microlensing event,
produced by a relatively massive MACHO.  The small variance can be
quantified. We determine a $\chi^2$ between the difference light curve
and a horizontal line (i.e. the hypothesis: no detectable
microlensing-induced changes within the measurement uncertainties)
defined by:
\begin{equation}
\chi^2=\frac{1}{N-1}\sum_{j=1}^{N} \frac{\Delta
m^2(t_j)}{\sigma^2(t_j)}.
\end{equation}
N is the number of data points of the difference light curve.  This
$\chi^2$-value measures the goodness of representing the data with
$\Delta m=0$. We obtain $\chi^2=1.0$ or $\chi^2=1.2$, depending on
which of the two light curve sets was interpolated (see above). This
means there is no statistically significant deviation between the
light curves of the two images.

However, in the difference light curve in Fig.~\ref{diff} there is a
small peak seen around day 750 (K97).  We tried in many ways to
establish the significance of these half dozen or so points which are
slightly but coherently above the ``zero''-line.  Interestingly, if
one goes back to the original individual light curves by K97, it can
be seen that at the Julian date corresponding to the peak in light
curve of image A, a similar peak was seen in the light curve of image
B. This could indicate a systematic effect during the observation.
Nevertheless, this ``peak'' in the difference light curve seems to
have been seen by Schild (\cite{Schild1996b}) as well with independent
data taken at a different observatory.

If real, such a peak in the difference light curve could be
interpreted as a short-duration, small-amplitude microlensing event in
image A.  In that case it could provide valuable information on the
mass of possible microlensing objects in the halo of the lensing
galaxy. Since we cannot prove with our data that this peak is real
(rather than a statistical fluke), in this paper we will argue more
conservatively that we do not detect any microlensing changes $\Delta
m_{\rm max} - \Delta m_{\rm min}$ higher than $0.05$ mag; we will
use this argument to put limits on the masses of the MACHOs in the
halo of the lensing galaxy.  This leaves open both the possibility
that this possible event is a fluke or that it is real. If the latter
turns out to be true, one can even draw some stronger quantitative
conclusions on the masses of possible MACHOs, rather than just exclude
certain regions in parameter space.

\subsection{Simulating microlensing light curves}

The exact shape of a microlensed quasar light curve depends on 
\begin{itemize}
\item the direction of the (projected) relative velocity vector
between lens and source
\item the (projected) relative positions of the MACHOs
\item the masses of the MACHOs 
\noindent and
\item the size of the continuum emitting region of the quasar. 
\end{itemize}

We do not and cannot know the exact positions of the MACHOs.
Hence we will not be able to predict or explain an individual
microlensed quasar light curve (in contrast to the ``low optical
depth'' regime of microlensing of stars in the Milky Way or Magellanic
Clouds). However, we can determine and analyse microlensed light curves
in a statistical sense; in particular, we investigate here the
distribution of total magnification variations.

We first determine the two-dimensional magnification variations due to
microlensing at different positions in the source plane with the
ray-shoo\-ting technique; we follow light rays backwards through an
arrangement of MACHOs randomly distributed in the plane of the lensing
galaxy, with surface mass density and shear as given by the lensing
model by Falco et al. (\cite{Falco1991}, priv. comm.) (see
Sect.~\ref{lensmodanddata}). The density of the deflected light rays
in the quasar plane corresponds to the relative magnification as a
function of position (Wambsganss \cite{Wambsganss1990a}). In these
simulations, we always use MACHOs with identical masses.\footnote{In
simulations with steep
	mass functions, e.g. Salpeter-like, most of the objects are
	near the lower cut-off; so the results of microlensing
	simulations with such mass functions are similar to those with
	all objects identical to the mean mass (Lewis \& Irwin
	\cite{Lewis1996}).}

We follow approximately of $10^{10}$ light rays and collect them in the
source plane in an array of 2500 by 2500 pixels.
We simulate microlensing light curves in the standard way (see,
e.g., Kayser \cite{Kayser1986} or Wambsganss \cite {Wambsganss1990a})
by evaluating the magnification along linear tracks across the
magnification patterns with a physical length L, equal to the length
the quasar traverses in the 160 day time span of the observed
difference light curve shown in Fig.~\ref{diff}. By assuming a
projected quasar velocity relative to the magnification pattern of
$v_{\rm t}=600$ kms$^{-1}$ (calculated using the method by Witt \& Mao
\cite{Witt1994}), this length is given by
\begin{equation}
{\rm L}=600\,{\rm km\,s}^{-1}\times 160\,{\rm days} = 8.3 \times
10^{14}\,{\rm cm}.
\end{equation}
We have neglected any motion of the MACHOs relative to each other
(Kundi\'c \& Wambsganss \cite{Kundic1993} and Wambsganss \& Kundi\'c
\cite{Wambsganss1995}).  Since velocities of stars in galaxies are in
general smaller than galaxy velocities, this effect cannot dominate
the bulk velocities. It merely slightly increases the value of the
transverse velocity of the quasar.

\subsection {MACHO mass and quasar size}

We calculated magnification patterns for both quasar images with
fractions of the halo mass contained in compact objects of $100\%$,
$50\%$ and $25\%$. For each of these mass fractions we produced
magnification patterns of varying physical side lengths of 20, 200 and
2000 Einstein radii. The Einstein radius in the source plane is defined as (e.g. Schneider
et al.
\cite{Schneider1992})
\begin{equation}
r_{\rm E}=\left( \frac{4{\rm G}M}{c^2} \frac{D_{\rm s}D_{\rm ds}}
{D_{\rm d}} \right)^{1/2}=3.7\times 10^{16}\, \sqrt {\frac {M}
{M_{\odot}}}\,h^{-1/2}\,{\rm cm}.
\end{equation}
From this expression one can see that a single magnification pattern
can be used to simulate light curves for various MACHO masses because
the physical length scales with the square root of the mass of the
MACHOs\footnote{An easy way to understand this ``scaling argument''
without invoking the concept of an ``Einstein radius'' is a follows:
Consider a certain region in the source plane.  Suppose that we know
the surface mass density of the MACHOs in the lens plane. Let us raise
the mass of all MACHOs by a factor~$q$.  If the surface mass density
is kept constant in the lens plane, this corresponds to blowing up the
length scale of the distribution of MACHOs by a
factor~$\sqrt{q}$. Since the deflection angle is proportional to the
mass and inversely proportional to the distance to the lens, the
magnification pattern in the source plane is thus also blown up by the
same factor of~$\sqrt{q}$.}; the fixed length the quasar traverses in
the source plane translates into different numbers of pixels for
different MACHO masses and pattern sizes.  The only limitation we have
for the investigation of various microlensing masses is the dynamic
range of the magnification pattern, which in our case is an array of
2500 by 2500 pixels. We used track lengths ranging from 10 pixels to
300 pixels, so that we could simulate three decades of MACHO masses
with one pattern. With the three different side lengths, however, we
were able to investigate the effects of MACHO masses $M$ ranging from
$10^{-8} M_{\odot}$ up to $10^{-1} M_{\odot}$ (in steps of factor
10). Due to the ``overlap'' we could check some masses on
magnification patterns with different side lengths and hence
cross-check the results.

\begin{figure*}
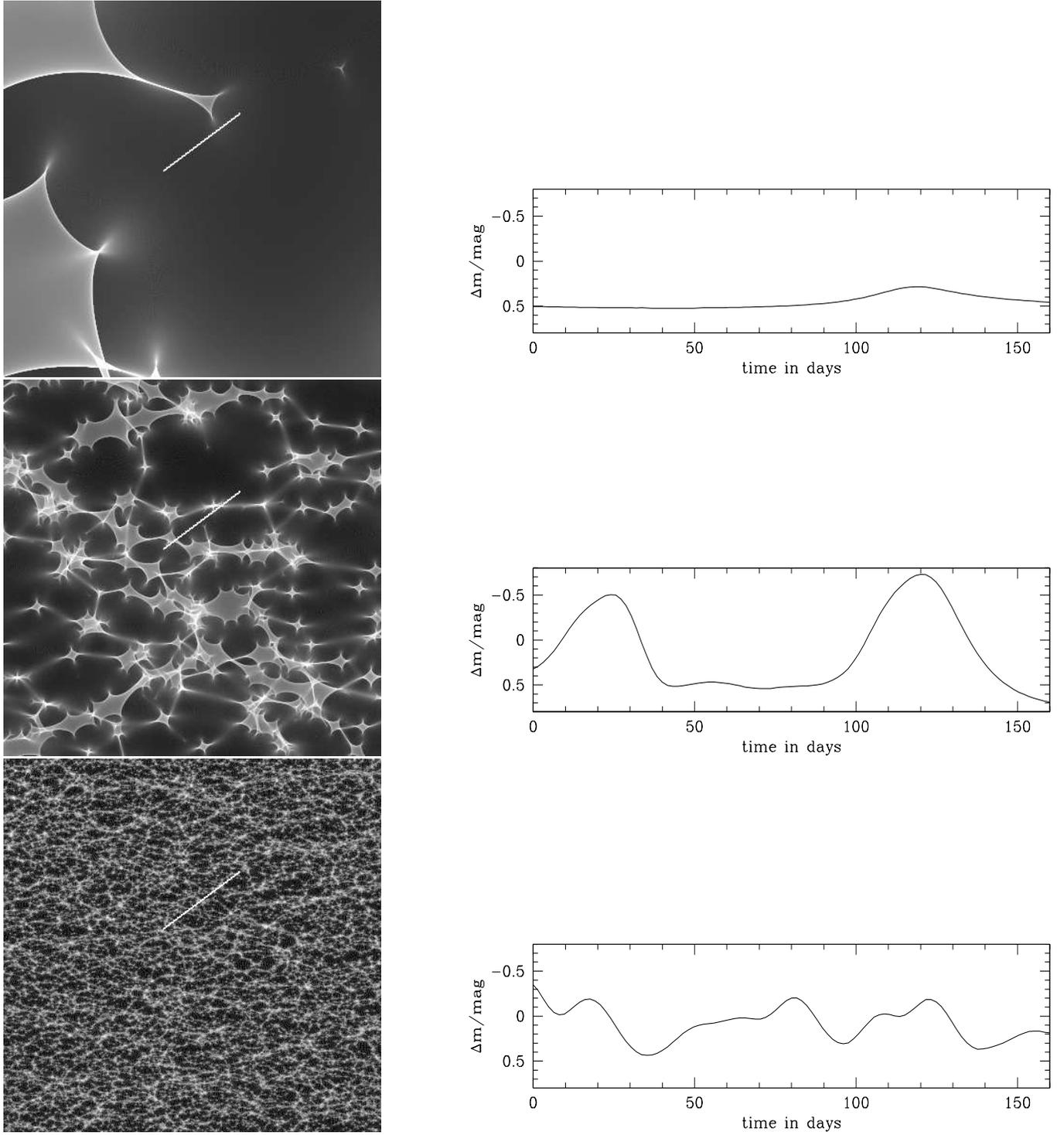

\resizebox{6.5cm}{!}{\includegraphics{h0863.f2a}}
\hfill
\resizebox{10.0cm}{3.25cm}{\includegraphics{h0863.f2b}}
\resizebox{6.5cm}{!}{\includegraphics{h0863.f2c}}
\hfill
\resizebox{10.0cm}{3.25cm}{\includegraphics{h0863.f2d}}
\resizebox{6.5cm}{!}{\includegraphics{h0863.f2e}}
\hfill
\resizebox{10.0cm}{3.25cm}{\includegraphics{h0863.f2f}}
\caption{Magnification patterns and light curves for different MACHO
masses. The three microlensing light curves on the right are simulated
by evaluating the magnifications along the white tracks in the three
magnification patterns on the left. The track length is chosen in such
a way that it corresponds to a 160 day light curve of Q0957+561 for an
assumed transverse velocity of the quasar across the magnification
pattern of $v_{\rm t}=600$ kms$^{-1}$. The magnification patterns are
greyscale-coded from white (high magnification) to black
(demagnification) and have side lengths of 4 Einstein radii (top), 40
Einstein radii (middle) and 400 Einstein radii (bottom), so that the
(fixed) track length corresponds to MACHO masses of $10^{-3}
M_{\odot}$ (top), $10^{-5} M_{\odot}$ (middle) and $10^{-7} M_{\odot}$
(bottom). The microlensing parameters of the magnifications pattern
are those obtained by Falco et al. (\cite{Falco1991}, priv. comm.) for
image A of Q0957+561; the surface density (in units of $\Sigma_{\rm
crit}$) is 0.18, and the shear is 0.32 (aligned horizontally in this
figure). The quasar is simulated with a Gaussian surface brightness
profile with a half-width of $3\times 10^{14}$ cm.}
\label{variability}
\end{figure*}

It follows from these considerations that the quasar traverses more
(fewer) characteristic lengths of the magnification pattern for
smaller (larger) MACHO masses during the observation period of 160
days. This implies that the light curve is more (less) variable for
smaller (larger) MACHO masses, so that one can derive limits on the
MACHO masses from the microlensing variability. This is illustrated
for three different mass scales in Fig.~\ref{variability}; the
variability of the microlensing light curves increases strongly with
decreasing MACHO mass. Qualitatively one can take from the
magnification patterns in Fig.~\ref{variability} that the probability
of observing no microlensing variation is practically zero for
$10^{-7} M_{\odot}$ MACHOs (bottom panel) whereas for the $10^{-3}
M_{\odot}$ MACHOs (top panel) this is not unlikely. These qualitative
statements will be quantified in the next section.

The size of the optical continuum region of the quasar, which we call
quasar size in the following, has an impact on the shape of a light
curve.  For extended objects the total magnification can be calculated
as a weighted mean of the magnifications at many points in the source
plane. In practice the effect of the source size can be accounted for
by convolving the two-dimensional magnification pattern with an
appropriate source profile.  Sharp features in the magnification
pattern -- especially the line-like caustics -- are thus smoothed out
by the brightness profile of the quasar.  The amplitude (smaller for
large sources) and the duration (longer for large sources) of
variations in a microlensing light curve hence depend on the source
size (see Wambsganss \& Paczy\'nski \cite{Wambsganss1991}).

We adopted a Gaussian profile for the surface brightness profile of
the quasar, where the source size is defined by the Gaussian width
$\sigma_{\rm Q}$. We used quasar source sizes ranging from
$\sigma_{\rm Q} = 10^{14}\,$cm up to $\sigma_{\rm Q}
=3\times10^{15}\,$cm (in steps of $\sqrt{10}$). These quasar
sizes are smaller than or of the order of one light day, which is an
upper limit on the quasar size that was obtained by Wambsganss et
al. (\cite{Wambsganss1990b}) and also Witt \& Mao (\cite{Witt1994})
for the quasar Q2237+0305. The case for even larger quasar sizes and
smaller MACHO masses was examined by Refsdal \& Stabell 
(\cite{Refsdal1991},\cite{Refsdal1993}) and Haugan (\cite{Haugan1996}).

We note that all our results for the MACHO masses and quasar sizes
may be scaled with $v=v_{\rm t}/600$ kms$^{-1}$ for various values
$v_{\rm t}$ of the transverse source velocity. MACHO masses scale
quadratically with $v$ because the scale of the magnification pattern
is proportional to the square root of the MACHO masses. Similarly,
the source sizes scale linearly with $v$.

\section{Results}

For each set of the two parameters MACHO mass and quasar size and for
each of the two quasar images A and B, we analysed 100\,000 randomly
chosen tracks across the magnification patterns. The same tracks were
used for different source sizes. For each light curve we determined
the difference between the highest and the lowest point of the light
curve $\Delta m_{\rm max}-\Delta m_{\rm min}$. We call this quantity
the total magnitude variation of the light curve.

\begin{figure*}
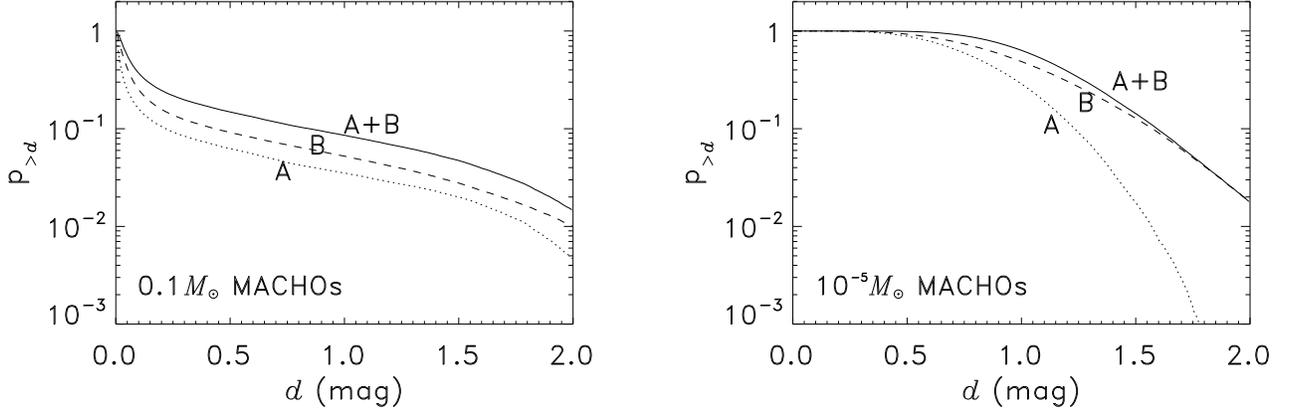

\resizebox{\hsize}{!}
{\includegraphics{h0863.f3a}\includegraphics{h0863.f3b}}
\caption{The probability $p_{>d}$ of observing a microlensing
variation greater than $d$ in the quasar Q0957+561 either in image A
(dotted line), image B (dashed line) or in at least one of the two
images (solid line). In the plot on the left the MACHO mass is $0.1
M_{\odot}$, whereas in the plot on the right it is $10^{-5}
M_{\odot}$. In both plots the quasar size is $10^{14}\,$cm and it is
assumed that the halo is completely made up of MACHOs.}
\label{distrib}
\end{figure*}

With these light curves, we can calculate the probability $p_{>d}$ of
observing a total variation greater than or equal to some value $d$
for each analysed parameter pair of MACHO mass and quasar size. As an
example, in Fig.~\ref{distrib} two integrated probability
distributions $p_{>d}$ are shown for MACHOs of mass $10^{-1}
M_{\odot}$ and $10^{-5} M_{\odot}$. In these plots it is assumed that
the quasar has a size of $10^{14}\,$cm and that the halo
mass is completely made up of MACHOs. Three lines are shown per plot;
the distributions for images A and B alone, as well as the joint
probability distribution where at least one quasar image has a total
variation greater than $d$. One can see that much stronger
microlensing variations are expected for small MACHO masses than for
large masses on these short time scales.

In the following, $p_{>d}$ refers only to the joint probability where
the microlensing variation is observed in at least one quasar image.
In order to compare the simulations with the observations, we
calculated the joint probability $p_{>0.05}$ for a grid of points in
our parameter space of MACHO masses and quasar sizes. We inferred in
Sect.~\ref{lensmodanddata} (Fig.~\ref{diff}) that the observations
indeed do not show a total variation greater than $0.05$~mag, so that
the quantity $p_{>0.05}$ can be viewed as the confidence level at
which we can exclude a particular parameter pair as not consistent
with the observations.

Various calculated values for $p_{>0.05}$ are given in
Tables~\ref{macho1} (for an assumed halo fraction of the MACHOs of
100\%), \ref{macho2} (for a halo fraction of 50\%), and \ref{macho3}
(for a halo fraction of 25\%). The table entries for parameter pairs
that are ruled out at a confidence level of $95\%$ and above are
highlighted in grey.  \definecolor{light}{gray}{0.8}
\begin{table}
\caption{Probabilities $p_{>0.05}$ (in percent) for measuring a total
microlensing variation greater than 0.05 mag in a 160 day difference
light curve of Q0957+561. In this table, it is assumed that MACHOs
constitute $100\%$ of the halo mass. The probabilities were calculated
using magnification patterns with three different side lengths (the
three main columns on the right) for several combinations of MACHO
mass and quasar size (indicated in the two columns on the left). No
values are given where the parameters were beyond the dynamical range
of the simulations. The statistical uncertainties are given in
brackets ($0.0$ is given where the uncertainty was below the rounding
precision). Probabilities above $95\%$ are highlighted in grey - the
respective parameter pairs are ruled out by the observations at the
95\% level.}
\label{macho1}
\begin{tabular}{@{}cclll@{}}
Macho & Quasar & \multicolumn{3}{c}{pattern side length}\\
mass & size & \multicolumn{3}{c}{(Einstein radii)}\\
($M_{\odot}$) & (cm) & 20 & 200 & 2000\\
\hline\\
$10^{-1}$ & $10^{14}$ & 58.5(1.6)\\
& $3\times10^{14}$ & 60.1(1.7)\\
& $10^{15}$ & 62.7(1.4)\\
& $3\times10^{15}$ & 55.8(1.0)\\
\\			
$10^{-2}$ & $10^{14}$ & 89.8(1.0)\\
& $3\times10^{14}$ & 90.9(0.9)\\
& $10^{15}$ & 89.8(0.7)\\
& $3\times10^{15}$ & 66.9(1.3)\\
\\
$10^{-3}$ & $10^{14}$ & \colorbox{light}{99.7(0.1)} & \colorbox{light}{99.8(0.0)}\\
& $3\times10^{14}$ & \colorbox{light}{99.7(0.1)} & \colorbox{light}{99.7(0.0)}\\
& $10^{15}$ & \colorbox{light}{97.1(0.6)} & \colorbox{light}{96.8(0.0)}\\
& $3\times10^{15}$ & 37.1(4.7) & 51.6(0.6)\\
\\
$10^{-4}$ & $10^{14}$ & \colorbox{light}{100.0(0.0)} & \colorbox{light}{100.0(0.0)}\\
& $3\times10^{14}$ & \colorbox{light}{100.0(0.0)} & \colorbox{light}{100.0(0.0)}\\ 
& $10^{15}$ & 89.6(1.2) & 93.8(0.2)\\
& $3\times10^{15}$ && 1.1(0.3)\\
\\
$10^{-5}$ & $10^{14}$ && \colorbox{light}{100.0(0.0)} & \colorbox{light}{100.0(0.0)}\\
& $3\times10^{14}$ && \colorbox{light}{100.0(0.0)} & \colorbox{light}{100.0(0.0)}\\
& $10^{15}$ && 58.8(2.5) & 60.6(0.2)\\
& $3\times10^{15}$ && 0.0(0.0) & 0.0(0.0)\\
\\
$10^{-6}$ & $10^{14}$ && \colorbox{light}{100.0(0.0)} & \colorbox{light}{100.0(0.0)}\\
& $3\times10^{14}$ && \colorbox{light}{98.0(0.4)} & \colorbox{light}{98.1(0.0)}\\
& $10^{15}$ && 2.4(0.3) & 3.2(0.2)\\
& $3\times10^{15}$ &&& 0.0(0.0)\\ 
\\
$10^{-7}$ & $10^{14}$ &&& \colorbox{light}{100.0(0.0)}\\
& $3\times10^{14}$ &&& 58.4(1.3)\\
& $10^{15}$ &&& 0.0(0.0)\\
& $3\times10^{15}$ &&& 0.0(0.0)\\
\\
$10^{-8}$ & $10^{14}$ &&& \colorbox{light}{98.4(0.2)}\\
& $3\times10^{14}$ &&& 0.0(0.0)\\
& $10^{15}$ &&& 0.0(0.0)\\
& $3\times10^{15}$ &&&
\end{tabular}
\end{table}
\begin{table}
\caption{Same as Table~\ref{macho1} for the case where $50\%$ of the
halo mass is contained in MACHOs.}
\label{macho2}
\begin{tabular}{@{}cclll@{}}
Macho & Quasar & \multicolumn{3}{c}{pattern side length}\\
mass & size & \multicolumn{3}{c}{(Einstein radii)}\\
($M_{\odot}$) & (cm) & 20 & 200 & 2000\\
\hline\\
$10^{-1}$ & $10^{14}$ &53.1(2.4)\\
& $3\times10^{14}$ & 51.9(0.9)\\
& $10^{15}$ &53.5(1.0)\\
& $3\times10^{15}$ & 50.1(0.8)\\
\\			
$10^{-2}$ & $10^{14}$ & 85.5(0.5)\\
& $3\times10^{14}$ & 86.1(0.6)\\
& $10^{15}$ & 86.0(0.5)\\
& $3\times10^{15}$ & 65.1(1.7)\\
\\			
$10^{-3}$ & $10^{14}$ & \colorbox{light}{98.9(0.1)} & \colorbox{light}{99.0(0.0)}\\
& $3\times10^{14}$ & \colorbox{light}{99.1(0.1)} & \colorbox{light}{98.9(0.0)}\\
& $10^{15}$ & \colorbox{light}{96.5(0.4)} & \colorbox{light}{96.0(0.0)}\\
& $3\times10^{15}$ & 39.7(9.3) & 36.1(0.8)\\
\\			
$10^{-4}$ & $10^{14}$ & \colorbox{light}{100.0(0.0)} & \colorbox{light}{100.0(0.0)}\\
& $3\times10^{14}$ & \colorbox{light}{100.0(0.0)} & \colorbox{light}{100.0(0.0)}\\
& $10^{15}$ & 90.9(2.1) & 89.5(0.2)\\
& $3\times10^{15}$ && 0.0(0.0)\\
\\			
$10^{-5}$ & $10^{14}$ && \colorbox{light}{100.0(0.0)} & \colorbox{light}{100.0(0.0)}\\
& $3\times10^{14}$ && \colorbox{light}{100.0(0.0)} & \colorbox{light}{100.0(0.0)}\\
& $10^{15}$ & & 40.2(2.6) & 39.4(0.2)\\
& $3\times10^{15}$ && 0.0(0.0) & 0.0(0.0)\\
\\			
$10^{-6}$ & $10^{14}$ && \colorbox{light}{100.0(0.0)} & \colorbox{light}{100.0(0.0)}\\
& $3\times10^{14}$ && 94.4(1.0) & 93.4(0.1)\\
& $10^{15}$ && 0.3(0.4) & 0.3(0.0)\\
& $3\times10^{15}$ &&& 0.0(0.0)\\
\\			
$10^{-7}$ & $10^{14}$ &&& \colorbox{light}{100.0(0.0)}\\
& $3\times10^{14}$ &&& 25.7(2.4)\\
& $10^{15}$ &&& 0.0(0.0)\\
& $3\times10^{15}$ &&& 0.0(0.0)\\
\\			
$10^{-8}$ & $10^{14}$ &&& 83.1(2.6)\\
& $3\times10^{14}$ &&& 0.0(0.0)\\
& $10^{15}$ &&& 0.0(0.0)\\
& $3\times10^{15}$ &&&
\end{tabular}
\end{table}
\begin{table}
\caption{Same as Table~\ref{macho1} for the case where $25\%$ of the
halo mass is contained in MACHOs.}
\label{macho3}
\begin{tabular}{@{}cclll@{}}
MACHO & Quasar & \multicolumn{3}{c}{pattern side length}\\
mass & size & \multicolumn{3}{c}{(Einstein radii)}\\
($M_{\odot}$) & (cm) & 20 & 200 & 2000\\
\hline\\
$10^{-1}$ & $10^{14}$ & 35.8(1.6)\\
& $3\times10^{14}$ & 36.2(1.6)\\
& $10^{15}$ & 37.1(1.8)\\
& $3\times10^{15}$ & 36.3(1.7)\\
\\			
$10^{-2}$ & $10^{14}$ & 74.1(1.1)\\
& $3\times10^{14}$ & 74.7(1.2)\\
& $10^{15}$ & 75.4(1.1)\\
& $3\times10^{15}$ & 54.9(2.6)\\
\\			
$10^{-3}$ & $10^{14}$ & \colorbox{light}{96.2(0.5)} & \colorbox{light}{95.9(0.1)}\\
& $3\times10^{14}$ & \colorbox{light}{96.5(0.5)} & \colorbox{light}{95.9(0.1)}\\
& $10^{15}$ & 92.1(1.3)	& 92.1(0.1)\\
& $3\times10^{15}$ & 16.0(8.2) & 20.2(0.3)\\
\\	
$10^{-4}$ & $10^{14}$ & \colorbox{light}{99.9(0.1)} & \colorbox{light}{99.9(0.0)}\\
& $3\times10^{14}$ & \colorbox{light}{99.8(0.1)} & \colorbox{light}{99.9(0.0)}\\
& $10^{15}$ & 73.8(7.8) & 81.5(0.2)\\
& $3\times10^{15}$ && 0.0(0.0)\\
\\			
$10^{-5}$ & $10^{14}$ && \colorbox{light}{100.0(0.0)} & \colorbox{light}{100.0(0.0)}\\
& $3\times10^{14}$ &&
\colorbox{light}{99.9(0.0)} & \colorbox{light}{99.8(0.0)}\\
& $10^{15}$ && 15.4(2.7) & 18.1(0.2)\\
& $3\times10^{15}$ && 0.0(0.0) & 0.0(0.0)\\
\\
$10^{-6}$ & $10^{14}$ & & \colorbox{light}{100.0(0.0)} & \colorbox{light}{100.0(0.0)}\\
& $3\times10^{14}$ & & 77.2(4.1) & 80.7(0.2)\\
& $10^{15}$ & & 0.0(0.0) & 0.0(0.0)\\
& $3\times10^{15}$ &&& 0.0(0.0)\\
\\			
$10^{-7}$ & $10^{14}$ &&& \colorbox{light}{100.0(0.0)}\\
& $3\times10^{14}$ &&& 7.2(1.2)\\
& $10^{15}$ &&& 0.0(0.0)\\
& $3\times10^{15}$ &&& 0.0(0.0)\\
\\
$10^{-8}$ & $10^{14}$ &&& 42.8(2.7)\\
& $3\times10^{14}$ &&& 0.0(0.0)\\
& $10^{15}$ &&& 0.0(0.0)\\
& $3\times10^{15}$ &&&
\end{tabular}
\end{table}
The uncertainties given in brackets are the standard deviations from
the mean of three different values for $p_{>0.05}$ that we derived
from three independent realizations of each magnification pattern
(using different random MACHO fields). The parts without entries
are regions in the parameter space which could not be accessed because
they were beyond the dynamical range of our simulations.

\begin{figure*}
\begin{center}\resizebox{12cm}{!}
{\includegraphics{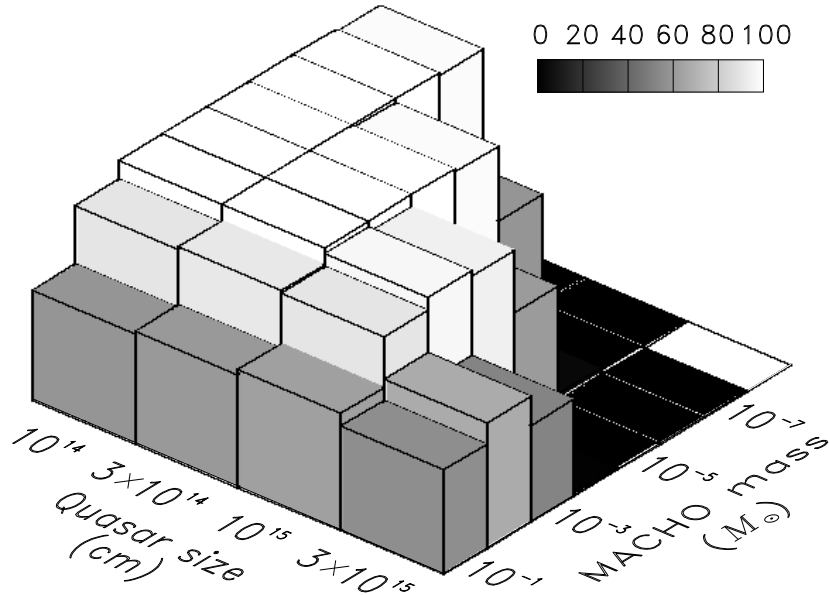}}
\end{center}
\caption{``Exclusion'' probability for certain MACHO
masses. Three-dimensional visualization of the probabilities
$p_{>0.05}$ (in percent) for measuring a total microlensing variation
greater than $0.05$ mag in a 160 day difference light curve of
Q0957+561 for a particular parameter pair of MACHO mass and quasar
size. The probabilities are indicated by the grey-shade of the bars
(see the key), the relative scale is visualized by the bar height.
The parameters of the blank field were beyond the dynamical range of
our simulations. These probabilities are those from Table~\ref{macho1}
for the largest available magnification pattern for each parameter
pair. It is assumed that MACHOs make up 100\% of the halo mass.}
\label{lego}
\end{figure*}

We have illustrated the results from Table~\ref{macho1} in
Fig.~\ref{lego}. In this plot the confidence levels are represented by
the height and the colour of the plotted bars.
The numbers from this plot and Tables~\ref{macho1}, \ref{macho2}
and~\ref{macho3} show that MACHO masses in the region from
$10^{-3}\,{\rm M}_{\odot}$ down to $10^{-5}\,{\rm M}_{\odot}$ can be
ruled out to make up a sizable fraction of the halo mass in the
lensing galaxy of Q0957+561 for quasar sizes smaller than $3\times
10^{14}\, $cm. In fact, for quasar sizes smaller than
$10^{14}\, $cm, we can exclude MACHO masses down to
$10^{-7}\,$M$_{\odot}$.

When looking at the uncertainties of the given probabilities in
Tables~\ref{macho1}, \ref{macho2} and~\ref{macho3} we find relatively
good agreement between the results from the three independent
realizations. Only for the light curves with pixel lengths of 300
pixels or 12\% of the side length of the magnification patterns
(corresponding to the smallest masses simulated with each
magnification pattern) the discrepancies between the probabilities get
larger. The reason is that the size of the light curve becomes
comparable to the whole field size and hence the light curves are not
strictly independent of each other. In order to get statistically
reliable results, one should use the $p_{>0.05}$ that was derived from
the largest magnification pattern (in terms of side lengths in units
of Einstein radii) that can be used to generate the light curve sample
in question. Incidentally, we find this way that some of the caustic
patterns show coherent structures on scales as large as 200 Einstein
radii.

\section{Discussion and conclusion}

In two years of observational data on the gravitationally lensed
double quasar Q0957+561 by Kundi\'c et al. (\cite{Kundic1997}) no
microlensing variation of the quasar larger than $0.05$ magnitudes was
observed. In our microlensing simulations we find that this rules out
a dominant population of compact objects in the halo of the lensing
galaxy from $10^{-3}\,M_{\odot}$ down to $10^{-5}\,M_{\odot}$ for
quasar sizes below $3\times 10^{14}\,{\rm cm}$. These
limits are almost independent of the fractions of the halo mass
contained in compact objects.

These results are consistent with the 2nd year results from the MACHO
microlensing search towards the Magellanic clouds (Alcock et
al. \cite{Alcock1997}). The most probable mass of MACHOs in the Milky
Way that emerges from their study lies around half a solar mass, which
is not excluded by our study. In fact, such large masses are not yet
probed by the observational data set we used (K97) because
microlensing effects by objects with masses of about one solar mass
are only becoming observable on time-scales of several years. In a
different data set on Q0957+561 (Schild \cite{Schild1996a}) there were
indications for some microlensing action which can be produced by
stars in this mass range. Very recently, Pelt et al. (\cite{Pelt1998})
analysed a large data set on Q0957+561. Similar to us, they cannot
find evidence for microlensing variations on short time scales.

Along the same lines, our results are consistent with the results of
Lewis \& Irwin (\cite{Lewis1996}) who found MACHO masses to be in the
range from $0.1 M_{\odot}$ up to $10.0$ $M_{\odot}$ from microlensing
in the quasar Q2237+0305.

The limits on compact objects in the halo of the lensing galaxy of
Q0957+561 will, however, improve with time since the monitoring of
Q0957+561 at the Apache Point Observatory (and also at other
observatories) is an ongoing project. If, for example, microlensing
variations above $0.05$ magnitudes are not observed for another
season, the limits from Tables~\ref{macho1}, \ref{macho2}
and~\ref{macho3} would become stronger and we could possibly rule out
compact objects with masses as high as $0.1 M_{\odot}$ or more.

In quasar microlensing, not only the mass of MACHOs, but also the size
of the quasar enters the calculations. With the current data, however,
we cannot constrain the quasar size very much. To do this, one would
need to find characteristic events in the difference light curve of
the two quasar images. If, for example, the peak in the difference
light curve (Schild \cite{Schild1996b}, K97) in Fig.~\ref{diff} around
day 750 is real, it would certainly be a valuable constraint for both
the masses of MACHOs and the quasar size.

Finally, in magnification patterns with side lengths of several
hundred Einstein radii we find coherent structures in the caustic
network on scales of 200 Einstein radii or more. In the lensing galaxy
of Q0957+561 these scales correspond to physical scales of the order
of $1.7 \sqrt {\frac{M}{M_{\odot}}}\,h^{-1/2}$pc for MACHOs of mass
$M$. On these scales the dark objects might not be distributed
completely randomly anymore. It would thus be interesting to
investigate further the effect of clustering of MACHOs on the microlensing
properties of galaxies.

\begin{acknowledgements}
We thank Rich Gott, Tomislav Kundi\'c, Avi Loeb, Bohdan Paczy\'nski,
Ue-Li Pen, Sjur Refsdal, Rudy Schild, Ed Turner and David Woods for
many helpful discussions and comments at various stages of this
project. This research was supported by the Deutsche
Forschungsgemeinschaft (DFG) under Gz. WA 1047/2-1.

\end{acknowledgements}

\end{document}